\documentclass[twocolumn,english,superscriptaddress]{revtex4-2}
\usepackage[T1]{fontenc}
\usepackage[latin9]{inputenc}
\usepackage{color}
\usepackage{units}
\usepackage{amsmath}
\usepackage{amssymb}
\usepackage{graphicx}

\makeatletter
\usepackage[colorlinks,citecolor=blue,
urlcolor=blue,hypertexnames=true,linkcolor=blue]
{hyperref} 

\usepackage{amssymb}
\usepackage{bbm}
\usepackage{scicite}

\makeatother

\usepackage{babel}
\begin{document}
\title{Distinguishability and \textquotedblleft which pathway\textquotedblright{}
information in multidimensional interferometric spectroscopy with
a single entangled photon-pair}
\author{Shahaf Asban}
\email{sasban@uci.edu}

\affiliation{Department of Chemistry and Physics \& Astronomy, University of California,
Irvine, California 92697-2025, USA}
\author{Shaul Mukamel}
\email{smukamel@uci.edu}

\affiliation{Department of Chemistry and Physics \& Astronomy, University of California,
Irvine, California 92697-2025, USA}
\begin{abstract}
Correlated photons inspire abundance of metrology-related platforms,
which benefit from quantum (anti-) correlations and outperform their
classical-light counterparts. While such demonstrations mainly focus
on entanglement, the role of photon exchange-phase and degree of distinguishability
have not been widely utilized in quantum-enhanced applications. Using
an interferometric setup we show that even at low degree entanglement,
when a two-photon wave-function is coupled to matter, it is encoded
with a reliable ``which pathway?'' information. An interferometric
exchange-phase-cycling protocol is developed, which enables phase-sensitive
discrimination between microscopic interaction histories (pathways).
We find that quantum-light interferometry facilitates utterly different
set of time-delay variables, which are unbound by uncertainty to the
inverse bandwidth of the wave-packet. We illustrate our findings on
an exciton model-system, and demonstrate how to probe intraband dephasing
in time-domain without temporal resolution at the detection. The exotic
scaling of multiphoton coincidence with respect to the applied intensity
is discussed.
\end{abstract}
\maketitle

\section{Introduction }

Interferometric spectroscopy introduces myriad of novel platforms,
aiming at revealing quantum information encoded on the wave-function
of multiple photons \citep{Raymer_2013,Lavoie_2020,Asban_2021,Dorfman_2021,Kushing_2020,Mukamel_2020}.
One of the most intriguing aspects of many-body quantum dynamics,
is the exchange statistics of indistinguishable particles. The wave-function
acquires a phase upon the exchange of two particles. This affects
their dynamics, and is detectable via unique interference patterns
in the correlations of two (or more) particles.Coincidence counting
of photons \citep{Mandel_1985}, current correlations of electrons
\citep{Bocquillon_2013} and fractional-charges (quantum-hall quasiparticles)
\citep{Chamon_1997,Deprez_2021} are notable examples. In quantum
electrodynamics, light-matter coupling can be represented as the sum
of all possible interaction histories (pathways). These pathways differ
by the order of the events, thus, multiphoton nonlinear processes
are potentially imprinted with their relative phases. Although the
exchange phase of photons is fixed and solely determined by their
bosonic nature, it can be effectively manipulated using a well established
interferometric setup for a pair of entangled photons \citep{Branning_1999}.
Here, we develop an exchange-phase-cycling scheme that scans through
different values of such a phase. We then, demonstrate the capacity
of multiphoton wave-function to encode and decipher matter information
inaccessible otherwise. Matter information-gain is physically manifested
in a reduced number of light-matter configurations, the ability to
switch between them, and a new set of time-delay variables with unique
characteristics. 
\begin{figure*}[t]
\noindent \begin{centering}
\includegraphics[scale=0.48]{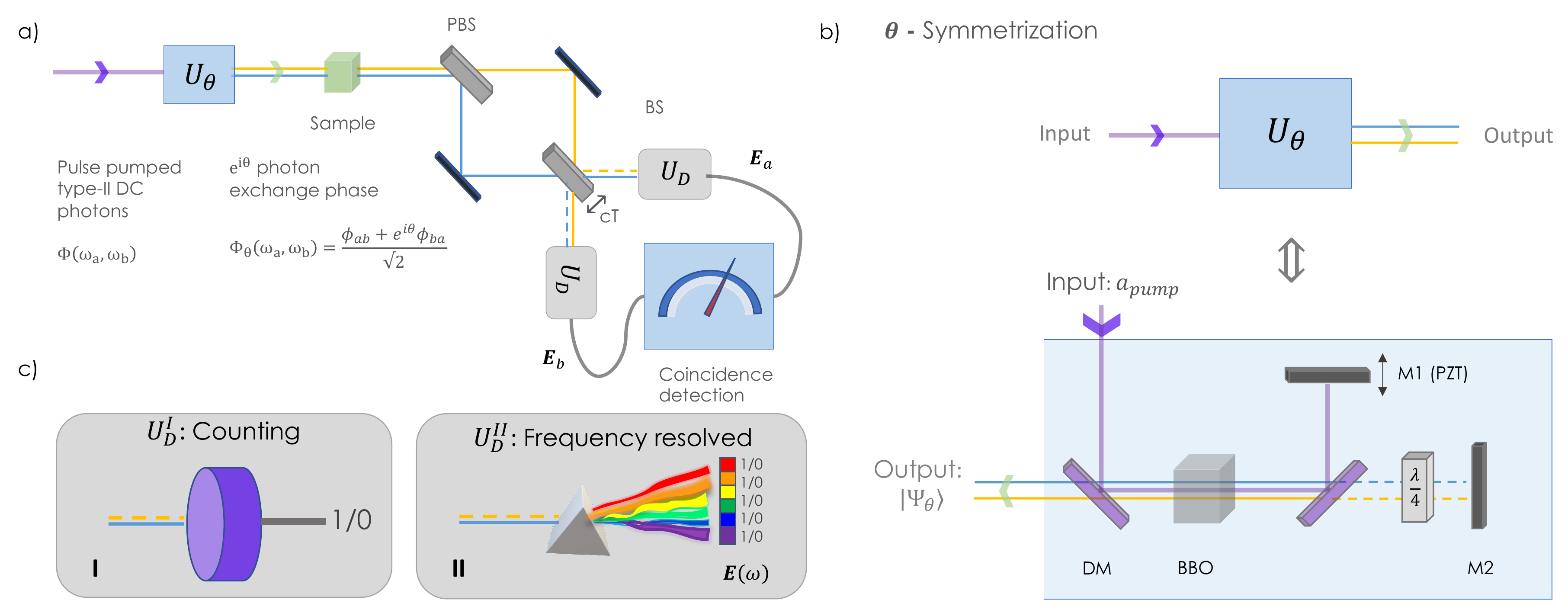}
\par\end{centering}
\caption{\textbf{Multidimensional interferometric spectroscopy setup}. (a)
Entangled photons with continuously variable exchange phase-$\theta$
described by $U_{\theta}$, interact with a sample. The pair is separated
by a polarization beam splitter (PBS), combined into a Hong-Ou-Mandel
(HOM) interferometer, and measured in coincidence following the detection
protocol $U_{D}$. The HOM setup introduces a relative temporal shift
to the photonic pathways, obtained by shifting the BS $T=\nicefrac{L}{c}$.
(b) The $\theta$-symmetrization denoted by $U_{\theta}$ is obtained
using a modified Michelson interferometer setup. The exchange-phase
$\theta$ grants a varying control over the photon-pair degree of
distinguishability. (c) Two detection protocols: (I) $U_{D}^{I}$
denotes the total photon counting signal (generating a $1/0$ event
registry), and (II) $U_{D}^{II}$ the frequency-resolved counting
(generating frequency-dependent $1/0$ list). \label{Fig 1}}
\end{figure*}

In this article, we study the multidimensional spectral information
generated by coupling an entangled photon-pair to matter, via combination
of interferometers depicted in Fig. $\text{\ref{Fig 1}}$. Our scheme
is composed of two interferometric setups: state preparation, followed
by 'reading' the quantum state encoded by light-matter information
exchange. This results in several notable differences in comparison
to the familiar semiclassical nonlinear optics. (i) While semiclassical
techniques scan time-delays between pulses, quantum interferometric
setups introduce new type of time-delay variables which are not conjugate
to the wave-packet bandwidth. (ii) Interferometric wave-mixing of
quantum light generates matter pathways unavailable classically due
to simultaneous detection of multiple photon propagation paths generated
at different times \citep{Asban_2021}. (iii) They allow separation
to new groups of pathways while delivering phase-sensitive read of
each process unmatched with classical light techniques. This allows
to reconstruct the temporal dynamics. (iv) The coincidence detection
obeys unique scaling relations between the applied intensity $I_{p}$,
the light-sample coupling and the detected signal. This enables to
avoid damaging disturbance to the sample, and eliminate unwanted signal
contributions (background). (v) Coincidence-counting singles out the
two-photon sub-space from the total signal, therefore, it restricts
the number of possible microscopic processes (selective). Consequently,
two-photon signals are sensitive to collective excitations (harmonic
and anharmonic), in contrast to single-photon counting which are generated
by matter anharmonicities \citep{Davydov_1964,Khalil_2001}.

Our proposed exchange-phase-cycling protocol projects the information
encoded in the multidimensional signal, onto lower dimension data
which reveal phase-dependent matter correlation functions. Moreover,
we harness the interferometric time-delay to probe the dephasing dynamics
of the sample, without resorting to time-resolved detection. The latter
is determined by the interferometer optical path-difference and is
not conjugate to the frequency measurement; paving the way to joint
time-frequency resolution beyond the Fourier limitation. 

\section{Results}

\subsection{The setup \label{sec:The-setup}}

The interferometric spectroscopy setup depicted in Fig. $\text{\ref{Fig 1}}$
is divided into preparation and detection stages. Both play an essential
role in acquisition of the nonlinear signal generated by the sample,
through its control parameters. In the preparation process, a modified
Michelson interferometer creates a photon-pair with tunable degree
of distinguishability, using the exchange-phase engineering described
below. At the detection stage, HOM interference is sensitive to the
post-coupling degree of distinguishability. Symmetric, antisymmetric
and asymmetric optical pathways, each carry valuable matter information
corresponding to different light-matter coupling history, enable temporal
reconstruction. At the end of this section, we summarize the control
parameters available using this setup.

\subsubsection{Preparation}

\begin{figure}

\begin{centering}
\includegraphics[scale=0.38]{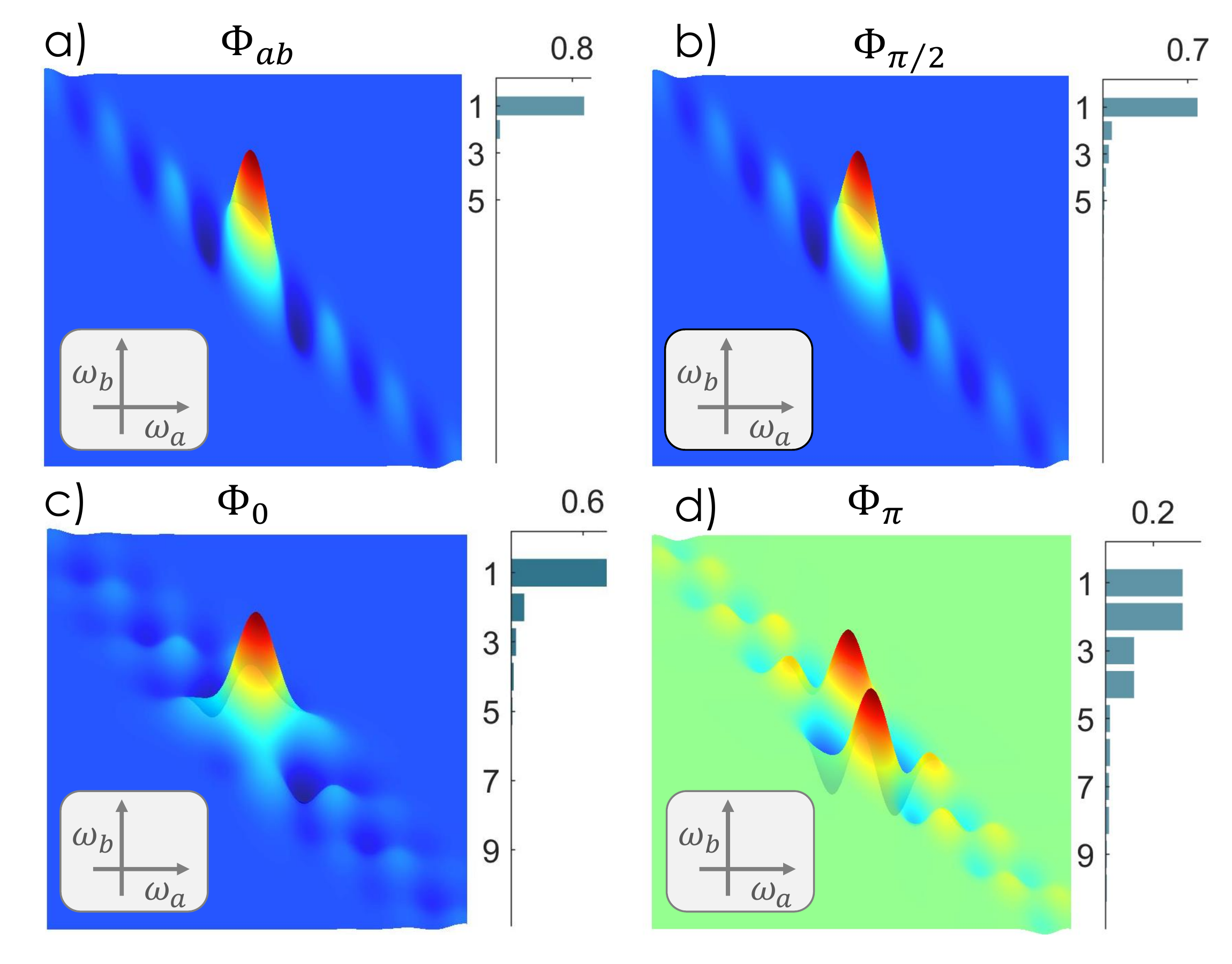}
\par\end{centering}
\caption{\textbf{Joint spectral amplitude.} The JSA in Eq. $\text{\ref{eq: JSA}}$
is presented for selection of $\theta$ values used on the exciton
model system. (a) The non-symmetrized amplitude two photon amplitude
in frequency domain, with Schmidt number $\kappa=1.01$. Symmetrized
amplitude using (b) $\theta=\frac{\pi}{2}$ with $\kappa_{\frac{\pi}{2}}=1.06$,
(c) $\theta=0$ with $\kappa_{0}=1.58$ and (d) antisymmetric $\theta=\pi$
with $\kappa_{\pi}=2.87$. Each bar-plots signify the Schmidt decomposition
resulting from the corresponding amplitude. \label{Fig 2}}
\end{figure}

A pump beam is directed into a modified Michelson interferometer using
a dichroic mirror, as shown in Fig. $\text{\ref{Fig 1}}$b and first
introduced in Ref. \citep{Branning_1999}. The beam passes through
a BBO ($\beta$ - barium borate) crystal, that generates a pair of
orthogonally (linearly) polarized entangled photons denoted as \emph{signal}
and \emph{idler. }The pair is fully characterized by the joint spectral
amplitude (JSA) $\phi\left(\omega_{a},\omega_{b}\right)$. The JSA
used in our calculations is given by $\phi\left(\omega_{a},\omega_{b}\right)={\cal A}_{p}\left(\omega_{a}+\omega_{b}\right)\varphi\left(\omega_{a},\omega_{b}\right)$
where ${\cal A}_{p}\left(\omega\right)=\exp\left[\nicefrac{\left(\omega-\omega_{p}\right)^{2}}{\sigma_{p}^{2}}\right]$
is a symmetric pump Gaussian envelope with bandwidth $\sigma_{p}$,
centered around $\omega_{p}$ \citep{Law_2000}. The phase-matching
factor $\varphi\left(\omega_{a},\omega_{b}\right)=\text{sinc}\left[\left(\omega_{a}-\bar{\omega}_{a}\right)T_{a}+\left(\omega_{b}-\bar{\omega}_{b}\right)T_{b}\right]$,
breaks the frequency exchange symmetry. Here $\bar{\omega}_{a/b}$
is the central frequency of the signal and idler, and $T_{a/b}=L\left(v_{a/b}^{-1}-v_{p}^{-1}\right)$,
where $L$ is the nonlinear crystal length and $v$ is the inverse
group velocity at the relevant central frequency $\left(\bar{\omega}_{a/b},\omega_{p}\right)$.
The JSA can exhibit strong exchange asymmetry, imprinting the horizontal
$\vert H\rangle$ and the vertical $\vert V\rangle$ polarization
quantum channels with distinct spectral signatures. All beams (pump,
signal and idler) are then split by a 50:50 BS, and each path is manipulated
separately. On one arm the polarizations are swapped by passing twice
through a $\nicefrac{\lambda}{4}$ plate, while the other undergoes
a controlled path delay introducing the phase $\theta$. In the second
passing through the BBO crystal (first during the generation), due
to the exchanged polarizations, the spectral profile is flipped from
$\phi\left(\omega_{a},\omega_{b}\right)$ to $\phi\left(\omega_{b},\omega_{a}\right)$.
The transmitted part of the beam from the DM is then given by the
$\theta$-symmetrized amplitude, resulting in the two-photon wave-function

\begin{subequations}

\begin{equation}
\vert\Psi_{\theta}\rangle=\int d\omega_{a}d\omega_{b}\Phi_{\theta}\left(\omega_{a},\omega_{b}\right)a^{\dagger}\left(\omega_{a}\right)b^{\dagger}\left(\omega_{b}\right)\vert\text{vac}\rangle\label{eq: 2-photon wavefunction}
\end{equation}

\begin{equation}
\Phi_{\theta}\left(\omega_{a},\omega_{b}\right)=\frac{1}{\sqrt{2}}\left[\phi\left(\omega_{a},\omega_{b}\right)+e^{i\theta}\phi\left(\omega_{b},\omega_{a}\right)\right],\label{eq: JSA}
\end{equation}

\end{subequations}

\noindent where $\vert\text{vac}\rangle$ is the noninteracting vacuum.
Broadband pumping of type-II parametric down-converter, is known to
generate photon-pairs with strong spectral distinguishing information
(see \citep{Grice_1997,Branning_1999}, and Sec. S1 of the SM). \textcolor{black}{We
shall show in Sec. $\text{\ref{subsec:Exchange-phase-cycling}}$ that
the asymmetric part of the JSA to $\left(\omega_{a},\omega_{b}\right)$
exchange plays a significant role in recovering the real part of the
matter correlation function for microscopic matter processes that
are symmetric to exchange. Fig. $\text{\ref{Fig 2}}$a depicts the
non-symmetrized JSA in frequency domain. Fig. $\text{\ref{Fig 2}}$b-d
then present $\theta=\nicefrac{\pi}{2},0$ and $\pi$ used in the
cycling protocols presented below. The JSA was computed with $\omega_{p}=4\:\text{eV}$,
with a $4\:\text{mm}$ nonlinear crystal with $T_{a}=6.1\,\text{fs}$
and $T_{b}=230\:\text{fs}$. The Schmidt number $\kappa_{\theta}\equiv\left[\sum_{n}p^{2}\left(n\vert\theta\right)\right]^{-1}$
computed by a diagnolization of the discretized single photon reduced
density matrix, following the procedure in Ref. \citep{Law_2000}.
Here, $p\left(n\vert\theta\right)$ is an eigenvalue that can be interpreted
as the $n^{\text{th}}$ mode probability using $\theta$-symmetrized
JSA. It provides a measure for the effective two-photon Hilbert space
\citep{Law_2000}. The Schmidt decomposition resulting from each JSA
is plotted in the bar-plot showing relatively low number of participating
modes, hence low degree of entanglement. Note that for $\theta=\pi$
Schmidt modes appear in pairs, as reported in Ref. \citep{Law_2000,Schlawin_2013b}.}

\subsubsection{Detection and coupling }

\noindent The coincidence signal is obtained by a two-photon expectation-value
at the detection plane. Since the detection event is represented by
an annihilation of a photon by the detector, it is useful to begin
the analysis from the detection stage. We then describe the light-matter
interaction stage using the joint density operator. Interferometric
setups can be simply described by a basis transformation between the
detection and interaction stages \citep{Asban_2021}. Two separate
coincidence detection protocols are considered; as depicted in Fig.
$\text{\ref{Fig 1}}$. The coincidence observables corresponding to
the two detection protocols $U_{D}^{I}$ and $U_{D}^{II}$ reflect
two-photon population in their respective domain (time-frequency).
Both setups contain several independent control parameters that manipulate
the excitation and detection mechanisms. Detection related control
parameters include the HOM relative delay $T$, and when resolved
$\left(U_{D}^{I}\right)$, the detected frequencies $\left(\omega_{a},\omega_{b}\right)$.
The exchange phase $\theta$, pump bandwidth $\sigma_{p}$, pump central
frequency $\omega_{p}$ and the pair central frequencies $\left(\bar{\omega}_{a},\bar{\omega}_{b}\right)$
constitute the excitation related control parameters. The first three
$\left(\theta,\sigma_{p},\omega_{p}\right)$ can be scanned continuously
in a single setup. Additional control parameters are the entangled
pair central frequencies $\left(\bar{\omega}_{a},\bar{\omega}_{b}\right)$,
which require a special quasi-phase-matching preparation procedure
\citep{Bahabad_2010}.

\paragraph*{Total coincidence signal -- $U_{D}^{I}$}

The total coincidence-count is described as annihilation of two modes
by two detectors. When the signal is not temporally resolved, the
annihilation time is integrated over. The corresponding observable
is given by the operator

\noindent {\small{}
\begin{multline}
\hat{{\cal O}}_{I}\left(t_{a},t_{b}\right)\\
=E_{a,R}^{\dagger}\left(\boldsymbol{r}_{a},t_{a}\right)E_{b,R}^{\dagger}\left(\boldsymbol{r}_{b},t_{b}\right)E_{b,L}\left(\boldsymbol{r}_{b},t_{b}\right)E_{a,L}\left(\boldsymbol{r}_{a},t_{a}\right).\label{eq: Counting Observable}
\end{multline}
}{\small\par}

\noindent Here, $E_{R}$ and $E_{L}$ are electric field superoperators,
corresponding to Hilbert-space operators that act from the right $E_{R}\rho\equiv E\rho$,
and left $E_{L}\rho\equiv\rho E$ of the density operator. The Hilbert-space
polarization-dependent field operator is given by $E_{\sigma}\left(\boldsymbol{r},t\right)=\sum_{\boldsymbol{k}}\sqrt{\frac{2\pi k}{\Omega_{Q}}}\hat{\epsilon}_{\sigma}\left(\boldsymbol{k}\right)a_{\boldsymbol{k},\sigma}\left(t\right)e^{i\boldsymbol{k}\cdot\boldsymbol{r}}$,
where $\hat{\epsilon}_{\sigma}\left(\boldsymbol{k}\right)$ is the
$\sigma$-polarization vector, $\Omega_{Q}$ is the quantization volume
($c=1$), and $a_{\boldsymbol{k},\sigma}\left(a_{\boldsymbol{k},\sigma}^{\dagger}\right)$
are annihilation (creation) operators obeying the bosonic commutation
relations $\left[a_{\boldsymbol{k},\sigma},a_{\boldsymbol{k}',\sigma'}^{\dagger}\right]=\delta_{\sigma,\sigma'}\delta_{\boldsymbol{k},\boldsymbol{k}'}$.
$\hat{{\cal O}}_{I}$ expresses annihilation of two-modes from the
left and right of the density operator, projecting the two-photon
subspace in the measurement. The coincidence signal is finally obtained
by taking the expectation value of Eq. $\text{\ref{eq: Counting Observable} }$
in the interaction picture

{\small{}
\begin{multline}
{\cal C}\left[\Lambda_{I}\right]=\int dt_{a}dt_{b}\\
\times\left\langle {\cal T}\hat{{\cal O}_{I}}\left(t_{a},t_{b}\right)\exp\left\{ -\frac{i}{\hbar}\underset{t_{0}}{\overset{t}{\int}}dsH{}_{\text{int},-}\left(s\right)\right\} \right\rangle ,\label{eq: Counting definition}
\end{multline}
}{\small\par}

\noindent where $\Lambda_{I}=\left\{ \omega_{p},\sigma_{p},\theta,T\right\} $
represents the set of control parameters available in this measurement
protocol. $\left\langle \hat{{\cal O}}\right\rangle \equiv\text{tr}\left\{ \hat{{\cal O}}\rho_{0}\right\} $
denotes the trace with respect to the initial state of the joint density
operator $\rho_{0}=\rho\left(t_{0}\right)$ and ${\cal T}$ is the
time ordering superoperator. The light-matter coupling is described
by the interaction superoperator, corresponding to the commutator
of the Hilbert-space interaction Hamiltonian and the density operator,
$H_{\text{int,\ensuremath{-}}}\rho\equiv\left[H_{\text{int}},\rho\right]$.
We adopt the the multipolar interaction Hamiltonian in the rotating
wave approximation (RWA) $H_{\text{int}}=\boldsymbol{E}^{\dagger}\cdot\boldsymbol{V}+H.c.$,
where $\boldsymbol{V}$ is the dipole lowering-operator. Within RWA,
emission of a photon is associated with energy decrease of the sample
while absorption with an increase. Note that the total coincidence
counting signal ${\cal C}_{I}$ is obtained by integration over photon
arrival times -- unresolved due to the lack of temporal gating.

\paragraph*{The frequency-resolved coincidence-counting -- $U_{D}^{II}$}

\noindent The frequency-resolved signal is obtained by a double annihilation
of optical modes, and defined by the corresponding frequency-domain
superoperators. These are related to time-domain counterparts using
Fourier transform $E_{\sigma}\left(\boldsymbol{r},t\right)=\int\frac{d\omega}{2\pi}\,e^{i\omega t}E_{\sigma}\left(\boldsymbol{r},\omega\right)$.
Similarly, the observable $\hat{{\cal O}}_{II}$ is given by

{\small{}
\begin{multline}
\hat{{\cal O}}_{II}\left(\omega_{a},\omega_{b}\right)\\
=E_{a,R}^{\dagger}\left(\boldsymbol{r}_{a},\omega_{a}\right)E_{b,R}^{\dagger}\left(\boldsymbol{r}_{b},\omega_{b}\right)E_{b,L}\left(\boldsymbol{r}_{b},\omega_{b}\right)E_{a,L}\left(\boldsymbol{r}_{a},\omega_{a}\right),\label{eq: Frequency observable}
\end{multline}
}{\small\par}

\noindent and the coincidence signal is obtained by the respective
expectation value 

{\small{}
\begin{multline}
{\cal C}\left[\Lambda_{II}\right]\\
=\left\langle {\cal T}\hat{{\cal O}}_{II}\left(\omega_{a},\omega_{b}\right)\exp\left\{ -\frac{i}{\hbar}\underset{t_{0}}{\overset{t}{\int}}dsH{}_{\text{int},-}\left(s\right)\right\} \right\rangle .\label{eq: Frequency counting definition}
\end{multline}
}{\small\par}

\noindent Here $\Lambda_{II}=\left\{ \omega_{a},\omega_{b},\omega_{p},\sigma_{p},\theta,T,\right\} $
are the corresponding control parameters. 

When implementing Eqs. $\text{\ref{eq: Counting definition} and \ref{eq: Frequency counting definition}}$,
it is crucial to note that the light-matter coupling are taken at
different stages of the interferometer. Consequently, they are given
in different basis sets and require the linear transformation described
next.

\paragraph*{Interferometric photon basis-transformation}

Due to the HOM interferometer, the optical modes involved in the light-matter
coupling, and the detected modes are given in different basis sets.
The transformation Jordan-Schwinger map) can be represented using
an $SU\left(2\right)$ rotation in the frequency-domain \citep{Yurke_1986,Jauche_1976,Mota_2004,Mota_2004_2,Mota_2016},
resulting in the input-output relation

\begin{equation}
\hat{{\cal R}}_{T}=\left(\begin{array}{cc}
t & ire^{i\omega T}\\
ire^{-i\omega T} & t
\end{array}\right).\label{HOM transf.}
\end{equation}

\noindent Here $t$ and $r$ are the transmission and reflection coefficients
obeying $\left|t\right|^{2}+\left|r\right|^{2}=1$, and $T$ is the
relative time-delay. For the 50:50 BS considered here, $t=r=\nicefrac{1}{\sqrt{2}}$.
The field in vector notation is given by $\boldsymbol{E}\left(\boldsymbol{r},\omega\right)=\left(E_{a}\left(\boldsymbol{r},\omega\right),E_{b}\left(\boldsymbol{r},\omega\right)\right)^{T}$,
under the HOM rotation the detected field is expressed by $\boldsymbol{E}\vert_{\text{detection}}\left(\boldsymbol{r},\omega\right)=\hat{{\cal R}}_{T}\boldsymbol{E}\vert_{\text{interaction}}\left(\boldsymbol{r},\omega\right)$.
In the following, we express all field operators in the basis set
of the interaction domain $\boldsymbol{E}\vert_{\text{interaction}}\equiv\boldsymbol{E}\left(\boldsymbol{r},\omega\right)$,
which requires the inverse rotation of the observable in Eq. $\text{\ref{eq: Counting Observable}}$
(see SI for detailed derivation) \citep{Asban_2021}. 

\begin{figure}
\begin{centering}
\includegraphics[scale=0.75]{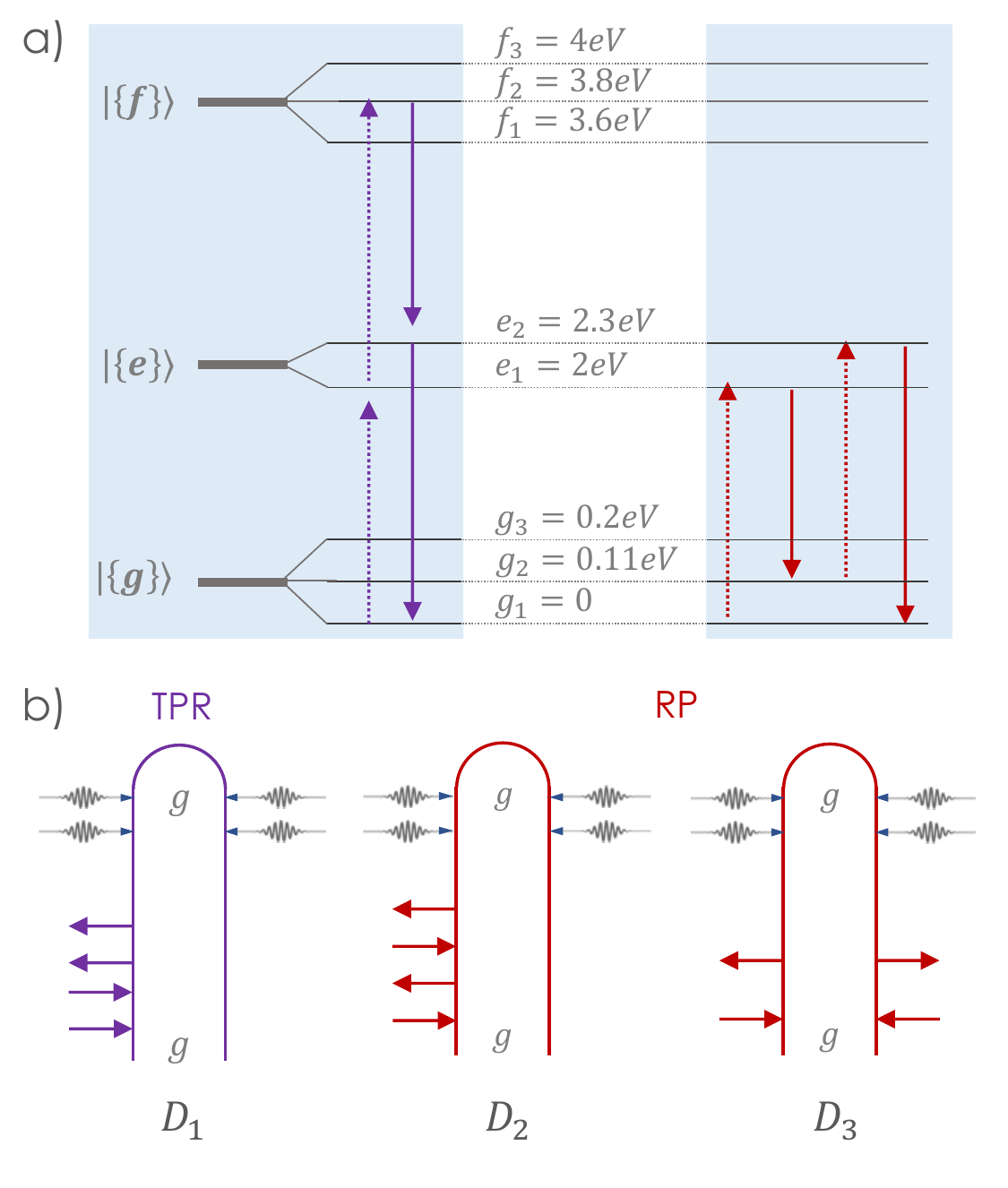}
\par\end{centering}
\caption{\textbf{Exciton model system and coupling pathways}. (a) Three-level
model system composed of ground $\vert\left\{ \boldsymbol{g}\right\} \rangle$,
single $\vert\left\{ \boldsymbol{e}\right\} \rangle$ and doubly excited
$\vert\left\{ \boldsymbol{f}\right\} \rangle$ manifolds. The (red)
arrows pointing interchangeably up and down correspond to Raman pathways.
The (purple) arrows arranged in two consecutive absorption followed
by two emissions correspond to two-photon resonances. (b) Diagrammatic
representation of the contributing microscopic light-matter processes.
Straight arrows represent an interaction corresponding to both fields.
Curved arrows represent the detection (annihilation) process. Diagram
$D_{1}$ involve double excitations (reaching the $f$-manifold) and
referred to as two-photon resonance (TPR). $D_{2}$ and $D_{3}$ describe
single excitations ($e$-manifold) and denoted Raman processes (RP).
The two distinct sub-groups of processes are depicted in panel (a).
\label{Fig 3}}
\end{figure}

\subsection{Application to an exciton model}

Consider the three-level exciton model-system depicted in Fig. $\text{\ref{Fig 3}a},$
with energy manifolds corresponding to the ground, singly and doubly
excited states $\left(g,e,f\right)$ respectively. We expand Eqs.
$\text{\ref{eq: Counting definition},\ref{eq: Frequency counting definition}}$
to $4^{th}$ order light-matter coupling $H_{\text{int}}$. In Fig.
$\text{\ref{Fig 3}}$b we show two groups of light-matter coupling
pathways; two-photon resonance (TPR) and Raman processes (RP), represented
in loop diagrams $D_{1}$, and $D_{2},D_{3}$ respectively. TPR interaction
sequences begin with two-photon absorption events $\vert\left\{ \boldsymbol{g}\right\} \rangle\rightarrow\vert\left\{ \boldsymbol{e}\right\} \rangle\rightarrow\vert\left\{ \boldsymbol{f}\right\} \rangle$,
while RPs involve $\vert\left\{ \boldsymbol{g}\right\} \rangle\leftrightarrow\vert\left\{ \boldsymbol{e}\right\} \rangle$
transitions (and do not involve the $f$ manifold). The diagrams describe
the light-matter interaction on a closed time-contour (Keldysh) in
which the ket (bra) evolve forward (backward) in time \citep{Mukamel_2008,Dorfman_2014}.
The sample is taken to be initially in the ground-state $\rho_{\mu}\left(-\infty\right)=\vert g_{1}\rangle\langle g_{1}\vert$,
each inward (outward) arrow denotes interaction-induced excitation
(de-excitation) of the sample, and the final state of the sample is
stated at the top of each diagram. Note that the reflection (interchanging
the bra-ket arrows) of all $D_{i}$ is obtained by complex conjugation.
We assume initially uncorrelated field-matter density operator $\rho\left(-\infty\right)=\rho_{\varphi}\left(-\infty\right)\otimes\rho_{\mu}\left(-\infty\right)$,
where $\rho_{\varphi}\left(-\infty\right)=\vert\Psi_{\theta}\rangle\langle\Psi_{\theta}\vert$
is the two-photon initial pure-state. 

Below, we show the signals obtained using the two coincidence-counting
detection protocols presented in Fig. $\text{\ref{Fig 1}}$, $U_{D}^{I/II}$.
The probability of observing each of the microscopic processes $\left(D_{i}\right)$
varies with the number of detected photons, consequently it is sensitive
to the final state of the sample; in contrast to a similar signal
obtained with a classical source. When single photons are detected,
all the processes contribute regardless to the number of generated
photons \citep{Schlawin_2013,Dorfman_2014b}. 

\begin{figure}
\begin{centering}
\includegraphics[scale=0.58]{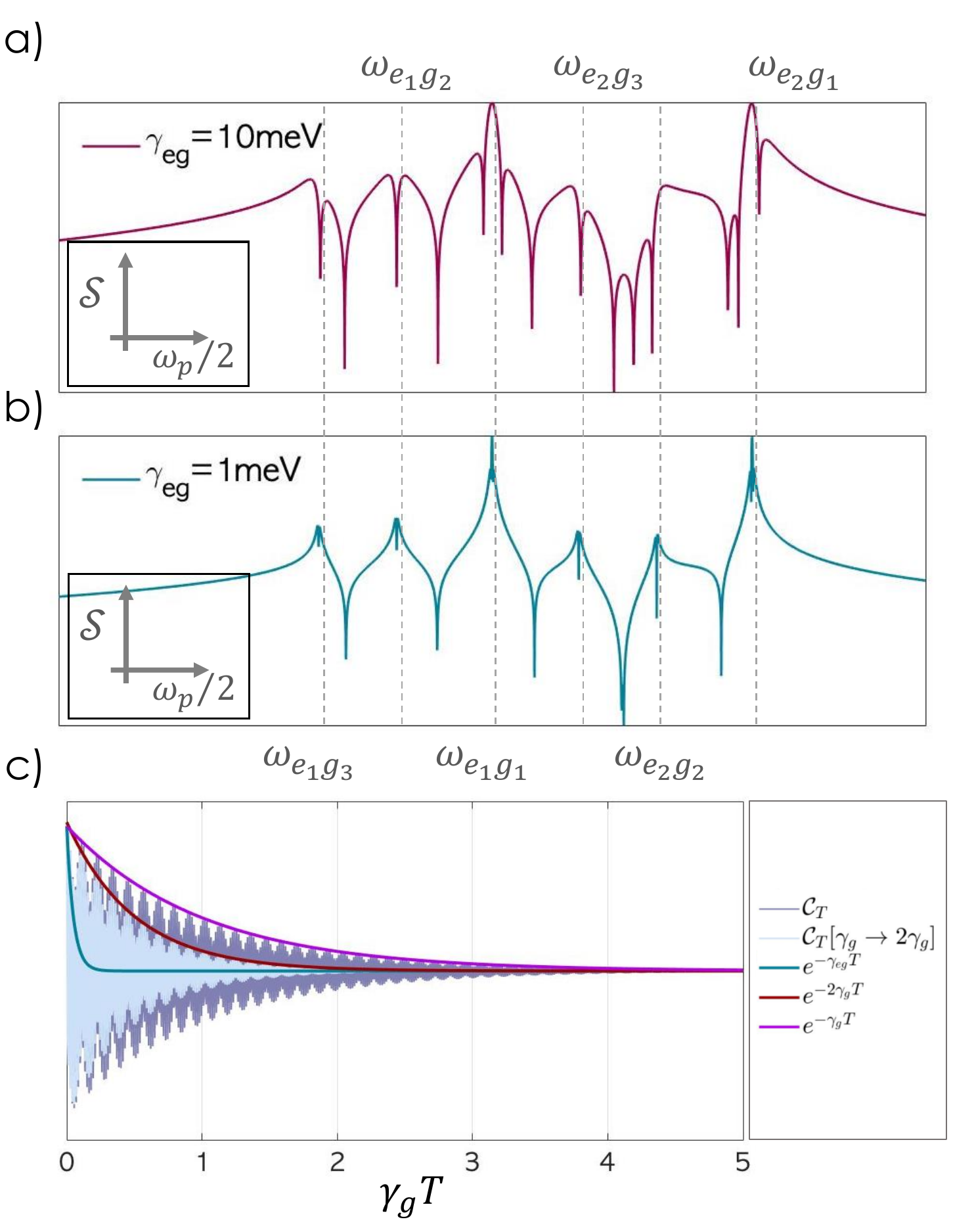}
\par\end{centering}
\caption{\textbf{Coincidence counting of the exciton system}. (a-b) logarithmic
plot of Eq. $\text{\ref{eq: Diphasing spectrum}}$, the spectra of
the total coincidence-counting ($U_{D}^{I}$ detection protocol) with
respect to the HOM delay $T$ and the pump central frequency $\nicefrac{\omega_{p}}{2}$.
The signal is obtained by a narrowband frequency-degenerate photon-pair,
corresponding to the intraband dephasing rate $\gamma_{g}=1\:\text{meV}$
(4 ps). Two interband rates are depicted (a) $\gamma_{eg}=10\,\text{meV}$
and (b) $\gamma_{eg}=1\,\text{meV}$. (c) Illustration of Eq. $\text{\ref{eq:time domain dephasing}},$
the total coincidence-counting signal as a function of the HOM delay
for two chosen dephasing rates $\gamma_{g}=1\,\text{meV}$ and $2\gamma_{g}$
and the respective exponential envelopes integrated over the pump
frequency. \label{Fig 4}}
\end{figure}

\subsubsection{Total-coincidence -- probing intraband dephasing }

The celebrated HOM dip, is an interference pattern of the total coincidence-count
of photon-pairs, obtained by varying their relative path-delay $T$
\citep{Hong_1987}. The count obtains its minimal value for $T=0$,
and vanishes altogether when the pair is completely indistinguishable.
Detection protocol $U_{D}^{I}$ manifests a HOM interference of the
pair posterior to the interaction with a sample. A signature of matter
energy-fluctuations shape the interference pattern in the presence
of the field. Under the conditioned elaborated below, it is possible
to isolate the contribution of diagram $D_{3}$ and probe the intraband
dephasing in time-domain. This measurement contains no time resolved
detection, using exclusively the HOM relative delay, which is not
conjugate to any frequency variable. 

The coincidence signal is derived for general pulse parameters and
corresponding symmetrization procedure (initial state) in Sec. S1
of the SM. The signal is derived using diagrams $D_{1},D_{2}$ and
$D_{3}$ resulting in Eq. S7. Simpler expressions are obtained for
a narrowband pump pulse, using degenerate phase-matching condition
for the entangled pair, fixing $\omega_{a}=\omega_{b}=\nicefrac{\omega_{p}}{2}$.
In this case, the phase-matching condition $\varphi\left(\omega_{a},\omega_{b}\right)$
is maximal for identical central frequencies $\bar{\omega}_{a}=\bar{\omega}_{b}$
of the entangled pair. We assume a narrowband pump of bandwidth (FWHM)
$\nicefrac{\Delta\lambda}{\lambda_{p}\leq10^{-2}}$ where $\lambda_{p}$
is the central frequency of the pump. The central frequency is scanned
in the range of $0.1-3\text{eV}$ attainable by pulse-duration in
the order of $\tau_{p}\approx10\,\text{ps}$. Combined with the degenerate
phase-matching factor the signal is maximal for $\omega_{a}=\omega_{b}=\nicefrac{\omega_{p}}{2}$.
Under these conditions, Eq. S7 reduces to, 

{\small{}
\begin{multline}
{\cal \,C}\left[\Lambda_{I}\right]={\cal C}\left[\omega_{p},\frac{\sigma_{p}}{\omega_{p}}\ll1,\theta=0,T\right]\\
\propto\mathfrak{Re}\text{tr}\left\{ VG^{\dagger}\left(\frac{\omega_{p}}{2}\right)V^{\dagger}\left[\mathbbm{1}-iG\left(T\right)\right]VG\left(\frac{\omega_{p}}{2}\right)V^{\dagger}\rho_{\mu}\left(-\infty\right)\right\} ,\label{eq: C_1}
\end{multline}
}{\small\par}

\noindent where we have used the fully symmetric initial state for
the field $\left(\theta=0\right)$, selectively isolating $D_{3}$.
Here $G\left(t\right)=-i\theta\left(t\right)e^{-iH_{\mu}t}$ is the
Green function of the sample and its Fourier transform $G\left(\omega\right)=\nicefrac{1}{\left(\omega-H_{\mu}+i\gamma\right)}$,
introducing the phenomenological dephasing rate $\gamma$ $\left(\hbar=1\right)$.
It is convenient to read Eq. $\text{\ref{eq: C_1}}$ from the density
matrix, the sample optically excited to a populated excited state
than de-excited back to the ground manifold where the observable $\mathbbm{1}-iG\left(T\right)$
is measured (see Eq. S9 of the SI for the full sum-over-states expression).
The time domain Green's function is evaluated at the ground state
(initial energy manifold), and thus reveal its temporal dynamics at
time $T$. Scanning $\omega_{p}$ and performing the Fourier transform
of the signal with respect to the pair $\left\{ \omega_{p},T\right\} $,
we obtain the spectra presented in Fig. $\text{\ref{Fig 4}a-b}$

\noindent 
\begin{equation}
{\cal S}=\int\frac{d\omega_{p}}{2\pi}{\cal \,C}\left[\Lambda_{I}\right]e^{i\nicefrac{\omega_{p}}{2}T}.\label{eq: Diphasing spectrum}
\end{equation}

\noindent Note that $T$ and $\omega_{p}$ are non-conjugate pair
and in principle can be resolved to arbitrary accuracy. Fig. $\text{\ref{Fig 4}a}$
depicts the spectrum obtained for intraband dephasing $\gamma_{g}=1\,\text{meV}$
$\left(4\text{ps}\right)$ and more rapid interband dephasing $\gamma_{eg}=10\,\text{meV}$
$\left(0.4\text{ps}\right)$. The latter contribute to the lineshape
broadening, limiting the frequency resolution. Fig. $\text{\ref{Fig 4}b}$
presents the same for equal interband and intraband dephasing $\gamma_{eg}=\gamma_{g}=1\,\text{meV}$
$\left(4\text{ps}\right)$. Fig. $\text{\ref{Fig 4}b}$ depicts the
spectrum obtained for equal interband and intraband dephasing $\gamma_{eg}=\gamma_{g}=1\,\text{meV}$
$\left(0.4\text{ps}\right)$. Fig. $\text{\ref{Fig 4}c}$ presents
the coincidence-count ${\cal C}\left[\Lambda_{I}\right]$, obtained
by varying the delay HOM delay parameter $T$ and tracing over $\omega_{p}$,

\begin{equation}
{\cal C}_{T}=\int\frac{d\omega_{p}}{2\pi}{\cal \,C}\left[\Lambda_{I}\right].\label{eq:time domain dephasing}
\end{equation}

\noindent We note that similar plot is obtained using monochromatic
$\omega_{p}$. Exponentially decaying envelope of the intraband dephasing
is visible (time-domain) is modulated by the fast oscillations at
the transition frequencies. Two possible dephasing rates are illustrated
$\gamma_{g}=1$ and $2\gamma_{g}=2\,\text{meV}$, as well as the fast
decay of $\gamma_{eg}$. 

\begin{figure*}
\begin{centering}
\includegraphics[scale=0.43]{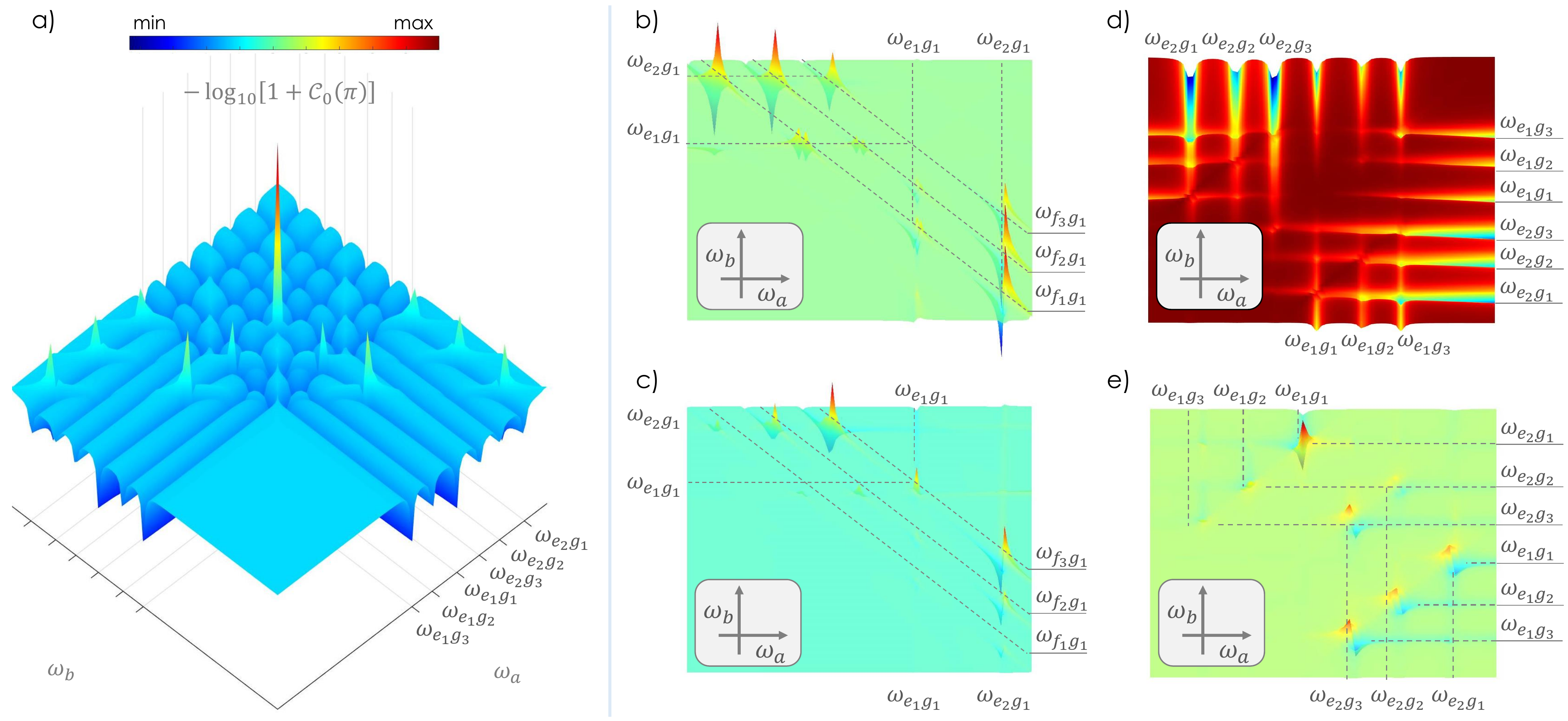}
\par\end{centering}
\caption{\textbf{Frequency-resolution of the exciton signal and exchange-phase-cycling}.
(a) 2D spectra obtained from Eq. $\text{\ref{S0}}$, using ultrafast
pump with $\sigma_{p}=0.9\text{meV}$ and scanning $\omega_{p}$ in
the range of $2-5\text{eV}$ on logarithmic scale. Exchange-phase-cycling
showing each of the components selectively. The imaginary (b) and
real (c) part of the TPR pathway given by the cycling protocols in
Eqs. $\text{\ref{S1}},\text{\ref{S2}}$ respectively. The dashed diagonal
lines follow the doubly excited transition $\omega_{a}+\omega_{b}=\omega_{f_{i}g_{1}}$.
The imaginary (d) and real (e) part of the RP pathway following the
cycling protocols in Eqs. $\text{\ref{S3}},\text{\ref{S4}}$. The
dashed lines denote transitions between the first excited and ground
manifolds $\omega_{e_{i}g_{j}}$. \label{Fig 5}}
\end{figure*}

\subsubsection{Frequency-resolved coincidence }

A natural extension of the coincidence-counting above, includes frequency
resolution of the detected photons depicted as detection protocol
$\left(U_{D}^{II}\right)$ in Fig. $\text{\ref{Fig 1}}$. The detected
photons add two dimensions to the above signal, resulting in a set
of control parameters $\Lambda_{II}=\left\{ \omega_{a},\omega_{b},\theta,T,\omega_{p},\sigma_{p}\right\} $.
Expansion of Eq. $\text{\ref{eq: Frequency counting definition}}$
to $4^{\text{th}}$ order, introduces another characteristic phase-factor
$\exp\left\{ i\eta\right\} $ due to the HOM delayed-path trajectories,
where $\eta=\left(\omega_{a}-\omega_{b}\right)T$ (see Sec. III of
the SM for full derivation). There are multiple ways to explore the
resulting high dimensional signal, one way is by fixing the pair $\left(\theta,\eta\right)$.
For brevity we use the shorthand notation ${\cal C}_{\theta}\left(\eta\right)\equiv{\cal C}\left[\Lambda_{II}\right]$
where $\eta$ can be fixed for any pair of frequencies $\left(\omega_{a},\omega_{b}\right)$
using the delay variable $T$. Also, it is convenient to introduce
the auxiliary functions corresponding to the real ${\cal R}_{i}\equiv\mathfrak{Re}\left\{ \Xi_{i}\right\} $
and imaginary ${\cal I}_{i}\equiv\mathfrak{Im}\left\{ \Xi_{i}\right\} $
parts of the respective pathway contributions. Here $i=\text{\ensuremath{\left\{  \text{TPR},\text{RP}\right\} } }$,
and $\text{\ensuremath{\Xi_{\text{i}}}}$ encapsulate all the microscopic
processes that contribute to each of the respective pathways (see
Sec. III of the SM for explicit expressions). Explicitly, the choice
$\left(\theta,\eta\right)=\left(0,\pi\right)$ results in 

\begin{align}
{\cal C}_{0}\left(\pi\right) & ={\cal I}_{\text{TPR}}\left(\omega_{a},\omega_{b}\right)+{\cal I}_{\text{RP}}\left(\omega_{a},\omega_{b}\right),\label{S0}
\end{align}

\noindent in which both pathway-groups are observed, as shown in $\text{\ref{Fig 5}}$a.
The calculations executed for an entangled pair generated by a broadband
pump with $\sigma_{p}=0.9\text{eV}$ ( $\approx1\text{fs}$ pulse).
The central frequency $\omega_{p}$ is scanned in the range of $2-5\text{eV}$
and all dephasing rates are identical $\gamma_{ij}=5\text{meV}$.
The Schmidt number $\kappa_{0}\approx2.7$ for the above parameters,
with a $L=0.4\,\text{mm}$. Fig $\text{\ref{Fig 5}}$a depicts all
contributing pathways, where we observe Fano-like resonances \citep{Limonov_2017}
located along the diagonal lines in which $\omega_{a}+\omega_{b}=\omega_{f_{i}g_{1}}$
such that $\omega_{a/b}=\omega_{e_{j}g_{1}}$ $\omega_{b/a}=\omega_{f_{i}e_{j}}$.
The RP pathways are observed along the lines corresponding to the
transitions $\omega_{e_{i}g_{j}}$. Similar results are obtained with
doubled bandwidth $\left(\sigma_{p}\right)$ corresponding to $\kappa_{0}\approx1.7$.
The coincidence-count vanishes for $\left(\theta,\eta\right)=\left(0,0\right)$
as we would expect for indistinguishable photons. 

\subsubsection{Exchange-phase-cycling \label{subsec:Exchange-phase-cycling}}

Pathway selectivity enables to study the dynamics in greater detail,
categorizing microscopic processes into distinguishable families related
by permutations of the light-matter interaction sequence. Frequency-resolved
detection $\left(U_{D}^{II}\right)$ depicted in Fig. $\text{\ref{Fig 1}}$
achieves just that -- isolating the real and imaginary part of each
process exclusively -- thanks to exchange-phase-cycling protocols
introduced below. 

We propose exchange-phase-cycling method whereby several signals with
different control parameters are combined to selectively observe desired
pathways. These exploit both interferometers; manipulating the effective
exchange phases prior and after the interaction respectively $\left(\theta,\eta\right)$.
The (modified) Michelson interferometer (Fig. $\text{\ref{Fig 1}}$b)
imprints any permutation of the initial photon-pair with a relative
phase-factor. The HOM detection interferometer introduces path-related
phase factor tot he detected photons. Certain combinations of $\left(\theta,\eta\right)$
are useful in reconstructing the real and imaginary parts of the signals
individually (See Sec. III of the SM for full derivation and final
expressions). For this purpose, it is useful to introduce the Fourier
transform of the coincidence signal ${\cal D}_{\theta}=\int dT\,e^{-i\Omega T}{\cal C}_{\theta}\left(\eta\right)$
(presenting 2D spectral map along the lines $\Omega=\omega_{a}-\omega_{b}$).
Combinations of ${\cal C}_{\theta}\left(\eta\right)$ and ${\cal D}_{\theta}$
render phase-sensitive reconstruction of the TPR and RP processes
possible exclusively. 

The cycling protocols are not unique, there are multiple choices of
linearly-dependent cycling protocols to achieve path selectivity.
Here, we display one cycling protocol resulting in the 2D spectra
presented in Fig. $\text{\ref{Fig 5}}$a-e. $\text{\ref{Fig 5}}$d-g
depicts the cycling protocol corresponding to the real ${\cal R}_{i}$
and imaginary ${\cal I}_{i}$ parts of the respective contribution

\begin{subequations}

{\footnotesize{}
\begin{align}
{\cal I}_{\text{TPR}}\left(\omega_{a},\omega_{b}\right) & ={\cal C}_{\pi}\left(\pi\right)\label{S1}\\
{\cal R}_{\text{TPR}}\left(\omega_{a},\omega_{b}\right) & ={\cal D}_{\frac{\pi}{2}}-{\cal D}_{-\frac{\pi}{2}}-{\cal C}_{\frac{\pi}{2}}\left(0\right)+{\cal C}_{-\frac{\pi}{2}}\left(0\right)\label{S2}\\
 & +{\cal C}_{\frac{\pi}{2}}\left(\frac{\pi}{2}\right)-{\cal C}_{-\frac{\pi}{2}}\left(\frac{\pi}{2}\right)\nonumber \\
{\cal I}_{\text{RP}}\left(\omega_{a},\omega_{b}\right) & ={\cal C}_{\frac{\pi}{2}}\left(0\right)+{\cal C}_{+\frac{\pi}{2}}\left(0\right)\label{S3}\\
{\cal R}_{\text{RP}}\left(\omega_{a},\omega_{b}\right) & ={\cal C}_{\frac{\pi}{2}}\left(0\right)-{\cal C}_{\frac{\pi}{2}}\left(0\right),\label{S4}
\end{align}
}{\footnotesize\par}

\end{subequations}

\noindent In our calculations we have used an entangled pair generated
by a pump with varying bandwidth; from a narrowband $\sigma_{p}=1.8\text{meV}$
corresponding to $\approx1\text{ps}$ pulse, to broadband $\sigma_{p}=0.9\text{eV}$
using a $\approx1\text{fs}$ pulse. The central frequency $\omega_{p}$
is scanned in the range of $2-5\text{eV}$ and all dephasing rates
are identical $\gamma_{ij}=5\text{meV}$. Due to the cycling parameters,
the real part of the response is strongly dependent on the degree
of asymmetry of the initial state. The asymmetric part of the JSA
gains stronger expression in the ultrafast regime, and become negligible
for a narrowband \citep{Branning_1999,Branning_2000}. Therefore,
the signal corresponding to the protocols in Eqs. $\text{\ref{S1}}-\ref{S4}$
are calculated using an ultrafast pump with $\sigma_{p}=0.9\text{meV}$
and depicted in Fig. $\text{\ref{Fig 5}}$b-e respectively.\textcolor{blue}{{}
}The Schmidt number $\kappa_{\theta}$ for the above parameters and
a $L=0.4\,\text{mm}$, varies between $1-5$ (depending on $\theta$).
Figs. $\text{\ref{Fig 5}}$b-c present the TPR pathway, where we observe
Fano-like resonances \citep{Limonov_2017} located along the diagonal
lines in which $\omega_{a}+\omega_{b}=\omega_{f_{i}g_{1}}$ such that
$\omega_{a/b}=\omega_{e_{j}g_{1}}$ $\omega_{b/a}=\omega_{f_{i}e_{j}}$.
Note that the transition $\omega_{f_{3}e_{1}}=\omega_{e_{1}g_{1}}=2\text{\,eV}$
is not resolved in ${\cal I}_{\text{TPR}}$ yet it appears in ${\cal R}_{\text{TPR}}$,
this stems from the antisymmetric nature of the JSA $\Phi_{\pi}\left(\omega_{a},\omega_{b}\right)$
which vanishes when $\omega_{a}=\omega_{b}$. While ${\cal I}_{\text{TPR}}$
is symmetric to the exchange, ${\cal R}_{\text{TPR}}$ is antisymmetric
and thus fully resolved. Nonetheless, different choices of cycling
may resolve this transition (see Sec. S5 of the SM and Fig. $\text{\ref{Fig 5}}$a
for such example). Figs. $\text{\ref{Fig 5}}$d-e depict the RP pathway,
scanning the single exciton energy manifold from which all transitions
$\omega_{e_{i}g_{j}}$ are visible. Finally, one can verify that the
information in panel (a) of Fig. $\text{\ref{Fig 5}}$ is composed
of a combination of panels (b) and (d), excluding the degenerate transitions
in which $\omega_{a}=\omega_{b}$. 

\subsection{Comparison with classical wave-mixing}

Nonlinear spectroscopic signals are usually described semiclassically
using a sequence of temporally-separated bright classical pulses (containing
many photons), that trigger matter dynamics and result in a single
photon \citep{Mukamel_1995}. The generated photon is modulated by
$m$ light-matter interactions, resulting in $m+1$ order correlation
function ($m+1$ wave-mixing). This nonlinear response is given by
the expectation value of the integrated change of the electric field
intensity $I=\int dt\,\dot{\left\langle {\cal I}\right\rangle }$
\citep{Marx_2008}. The intensity is a single photon quantity, related
to the electric field operator ${\cal I}=\boldsymbol{E}^{\dagger}\boldsymbol{E}$.
Quantum sources can reach superior signal-to-noise ratio scaling (Heisenberg
limit) \citep{Helstrom_1976,Giovannetti_2004,Giovannetti_review_2011,Napolitano_2010,Napolitano_2011},
enabling reduced radiation exposure for comparable measurement certainty.
Particularly, to improve the resolution of sensitive samples, limited
by radiation dose constraints \citep{HOWELLS_2009}. One way to benefit
from the quantum properties of the EM field, is via direct coupling
of quantum-light with to the sample (e.g. entangled photons, squeezed
states). Sample stimuli using such sources has shown to yield remarkable
control over population dynamics and pathway selection \citep{Schlawin_2013,Roslyak_2009,Dorfman_rev_2016}.
Alternatively, as done here, one can probe quantum effects of the
emitted radiation directly via multiple photons counting (e.g. antibunching
\citep{Kimble_1977}, superesolved imaging \citep{Hell_1994,Mouradian_2011,Schwartz2013,Tenne_2019}). 

\begin{figure}
\begin{centering}
\includegraphics[scale=0.65]{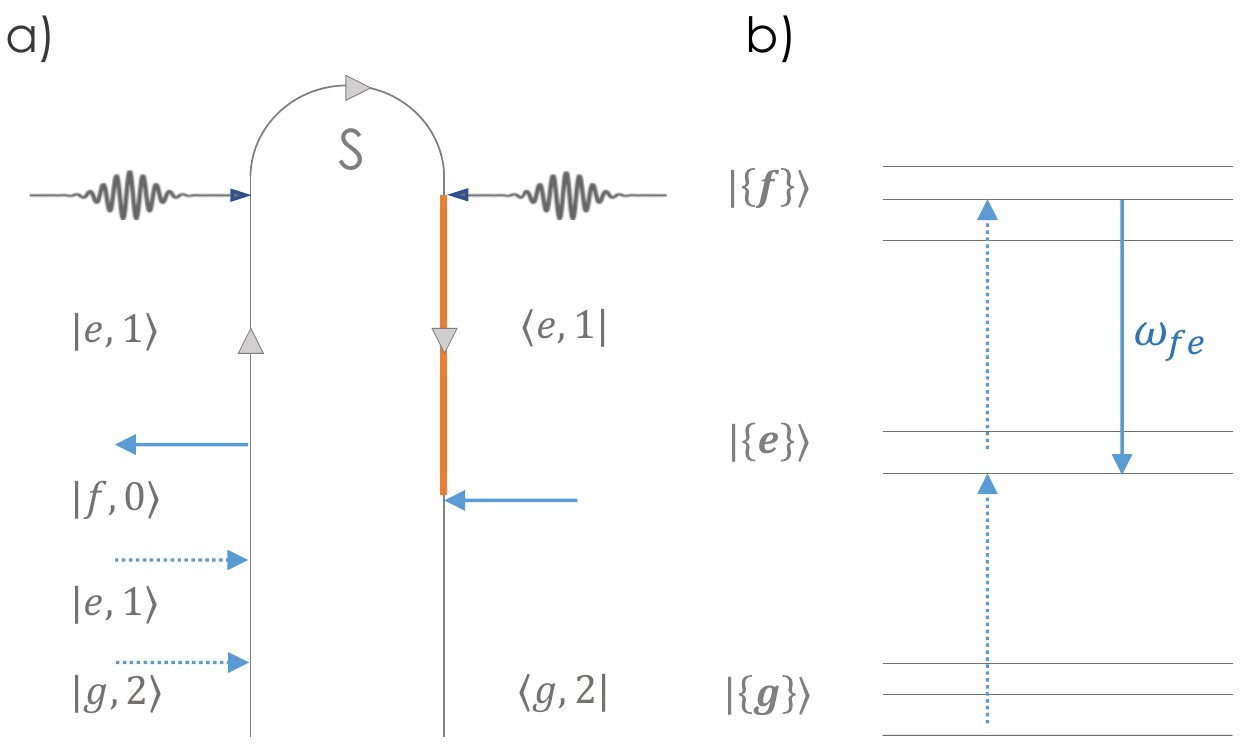}
\par\end{centering}
\caption{\textbf{TPR process} \textbf{missing in the coincidence signal}. (a)
A single photon TPR process labeled $S$, that involve double excitations
($f$-manifold). The joint field-matter is specified after each interaction
event on the diagram. The highlighted (orange) line correspond to
the backwards propagator responsible for the elimination of the (harmonic)
collective excitations in single photon nonlinear signals. (b) Schematic
representation of the process. \label{Fig 6}}
\end{figure}

Following this reasoning, we are interested in multiple photon-detection.
The resulting wave-mixing is denoted $\left(n+m\right)$-WM corresponding
to the application of $m$ fields and detection of the $n$ photons.
While $\left(m+1\right)$-WM depend on several pathways, $\left(n+m\right)$-WM
are naturally restricted to lower number of diagrams and may not be
written in the form of amplitude-square. Specifically to the above
setup, coincidence counting eliminates the single photon diagram depicted
in Fig. $\text{\ref{Fig 6}}$ denoted $S$. This occurs since only
one photon is populated in final state. This cancellation yields significant
change in the observed physics compared to single-photon (intensity)
signal. Intensity observables generated from nonlinear optical processes,
are generated from anharmonicities in matter. Thus, collective excitations
in the $f$ manifold with energies $\epsilon_{f_{k}g}=\epsilon_{e_{i}g}+\epsilon_{e_{k}g}$
vanish. This stems from the opposite evolution of the last interaction
comparing $D_{1}$ in Fig. $\text{\ref{Fig 3}}$ with $S$ in Fig.
$\text{\ref{Fig 6}}$. For small dephasing rates the forward and backward
resonances inherit opposite relative sign, verified using the Sokhotski-Plemelj
theorem $\lim_{\gamma\rightarrow0}\nicefrac{1}{\omega\pm i\gamma}=\mp i\pi\delta\left(\omega\right)+\text{pp}\left(\frac{1}{\omega}\right)$
where $\text{pp}$ denotes the \emph{principle part}. Due to the elimination
of $S$, the coincidence signal here is sensitive to collective excitations
in TPR processes.

$\left(n+m\right)$ WM also give rise to different intensity-coupling-signal
scaling relations. For example, double excitation signals induced
by entangled pairs are known to scale linearly (rather than quadratically)
with the pump intensity $\propto I_{\text{p}}$ \citep{Javanainen_1990,Lee_2006,Guzman_2010}.
This unique effect permits studying doubly excited manifold with smaller
probability of ground-state bleaching, thus potentially reduce sample
damage. While single photon detection events scale linearly with the
pump $\propto I_{\text{p}},$ the two-photon signal scales quadratically
$\propto I_{\text{p}}^{2}$ maintaining the double excitation probability
linear with the pump. This allows application of lower intensities
per desired detection gain, improving signal to noise ratio \citep{Lantz_2014,Bolduc_2017,Reichert_2018}.
This principle can be generalized to $n$ photon population detection
in a straightforward manner. 

\section{Discussion}

The measurement of a fixed number of photons using coincidence detection
narrows down the observed physics of the sample. Only microscopic
pathways which are terminated with a certain final-state contribute.
In contrast, nonlinear signals obtained with classical light are not
sensitive to the last state of the matter. Specifically, two-photon
coincidence of weakly-coupled entangled pair to a sample generates
a unique (2+2) wave-mixing, eliminating some microscopic pathways.
Like four-wave-mixing (4-WM), the signal depends on a four point dipole
correlation function of matter. The difference is that the elaborate
detection does not single out one field, as the 4-WM signal, but two
fields are detected. We can thus view the process as a generalized
4-WM. To Avoid confusion we simply refer to it as (2+2) WM, corresponding
to the number of applied and detected fields.

In the first detection scheme, we have considered the total photon
coincidence-count in the absence of spectral and temporal resolutions.
The two-natural control-parameters of the setup depicted in Fig. $\text{\ref{Fig 1}}$
are the HOM relative delay $T$, and the central frequency of the
pair-generating pump $\omega_{p}$. A fascinating effect occurs when
considering a degenerate phase-matching condition with a symmetric
narrowband JSA $\left(\theta=0\right)$; such that the photon-pair
are distributed sharply around half the pump frequency $\nicefrac{\omega_{p}}{2}$.
The TPR pathways vanish and the only RP pathway that survives is diagram
$D_{3}$ (Fig. $\text{\ref{Fig 3}}$). This generates the following
dynamics: (1) the density matrix of the sample is excited then de-excited
back to the initial band (here, ground) (2) Green's function of the
system at time $T$ is measured. Since $T$ is controlled with arbitrary
precision, the intraband dephasing dynamics can be reconstructed temporally
as shown in Fig. $\text{\ref{Fig 4}}$a-c. This provides a potential
platform to characterize the system's coupling to its environment.
Small $T$ expansion can reveal moments of the\textcolor{blue}{{} }sample
Hamiltonian $\left\langle H^{n}\right\rangle $ and provided in Sec.
S2 of the SM. This provides a compelling direction for future study
aligned with the great efforts invested in studying decoherence and
energy leaks in current quantum technologies.

The second detection protocol, involves frequency-resolve coincidence-counting.
By cycling values of $T$ and the entangled pair exchange-phase $\theta$
we are able to discriminate between TPR and RP pathways. Known pathway-selection
protocols typically generate destructive interference to suppress
certain populations \citep{Schlawin_2013}, or induce distinct scaling
of each process with the intensity \citep{Roslyak_2009}. Here, we
select pathways by projecting the high-dimensional signal to process-specific
data, without reducing the event probability of other processes. Moreover,
by combining signals with different $\theta$ and $T\left(\omega_{a}-\omega_{b}\right)$
(phase cycling), we demonstrate the ability to obtain the phase of
four-point matter correlation-function in Fig. $\text{\ref{Fig 5}}$b-e.
This implies that temporal reconstruction is possible by straight
forward Fourier transform. Interestingly, the cycling protocols related
to the real part of the correlator require an asymmetric JSA for exchange,
inherited from the pump in the ultrafast pumping regime. This suggests
that obtaining temporal behavior of the correlator necessitates an
ultrashort generating pump -- which is compatible with temporal resolution
-- despite the that the measurement is conducted in frequency domain.
An ultrashort pump generate a spontaneous pair that carry an identity-revealing
spectral information \citep{Grice_1997}. \textcolor{black}{The resulting
distinguishability renders a ``which pathway?'' information available.}\textcolor{blue}{{}
}The relation between the JSA asymmetry and the temporal resolution
upper-bound merit further study, since the temporal and frequency
control parameters are not conjugate quantities. Notably, these effects
occur specifically at low entanglement values and steered by the variable
effective exchange phase. 

In addition to the benefits listed above, matter-induced field nonlinearities
may also improve the frequency resolution \citep{Napolitano_2010,Napolitano_2011}.
Correlation-based detection techniques -- such as coincidence --
can reveal such electromagnetic field nonlinearities. These correlations
are signatures matter-induced photon-photon coupling \citep{Asban_2019},
free-electrons coupling \citep{Asban_2021b} and thus carry matter
information that is imprinted in the post-interaction counting-statistics.
Characterization of the 'reading' process that access this information,
is one of the central goals of interferometric-spectroscopy. 

\section{Materials and methods}

\subsubsection*{Schmidt number calculation}

The Schmidt number is a function of the (single photon) reduced density
matrix. It enables to represent the JSA in the following way

{\small{}
\begin{equation}
\Phi_{\theta}\left(\omega_{a},\omega_{b}\right)=\sum_{n}\sqrt{p\left(n\vert\theta\right)}\psi_{n}\left(\omega_{a}\right)\phi_{n}\left(\omega_{b}\right).
\end{equation}
}{\small\par}

\noindent Here $\left\{ \psi_{n},\phi_{n}\right\} $ are single photon
eigenfunctions and their weights $p\left(n\vert\theta\right)$. To
obtain this representation we solve $p\left(n\vert\theta\right)\psi_{n}\left(\omega\right)=\int dk'\,K_{1}\left(\omega,\omega'\right)\psi_{n}\left(\omega'\right)$
and $p\left(n\vert\theta\right)\phi_{n}\left(\omega\right)=\int dk'\,K_{2}\left(\omega,\omega'\right)\phi_{n}\left(\omega'\right)$.
These kernels are found from the reductions $K_{1}\left(\omega,\omega^{\prime}\right)=\int d\omega^{\prime\prime}\,\Phi_{\theta}\left(\omega,\omega^{\prime\prime}\right)\Phi_{\theta}^{*}\left(\omega^{\prime},\omega^{\prime\prime}\right)$
tracing the second frequency variable, and $K_{2}\left(\omega,\omega^{\prime}\right)=\int d\omega^{\prime\prime}\,\Phi_{\theta}\left(\omega^{\prime\prime},\omega\right)\Phi_{\theta}^{*}\left(\omega^{\prime\prime},\omega^{\prime}\right)$.
They can be interpreted as single-photon correlation functions \citep{Law_2000}.
To obtain the Schmidt spectrum and characterize the degree of entanglement,
we discretize the kernels and numerically solve the integral eigenvalue
equations. We have used a $900\times900$ kernel discretization grid
and calculated separately for each $\theta$. The Schmidt number which
is a measure of the effective Hilbert space dimensionality is obtained
by 

\begin{equation}
\kappa_{\theta}\equiv\frac{1}{\sum_{n}p^{2}\left(n\vert\theta\right)}.
\end{equation}

\subsubsection*{Intraband dephasing signal computation}

Eq. $\text{\ref{eq: C_1}}$ reveals the intraband dephasing in time-domain
using solely the HOM relative delay $T$ as scanning time-variable.
We express the correlation functions using the dipole lowering operator
$V=\sum_{i>j}\mu_{ij}\vert j\rangle\langle i\vert$ and its complex
conjugate, where $\left(i,j\right)$ label energy eigenstates of the
exciton system corresponding to $\vert g_{i}\rangle,\vert e_{i}\rangle$
and $\vert f_{i}\rangle$. The expectation value in Eq. $\text{\ref{eq: C_1}}$
is requires Green's function of the sample in frequency domain $G\left(\omega\right)=\left(\omega-H+i\epsilon\right)^{-1}$,
where $H$ is the Hamiltonian of the sample. With these definitions,
we obtain the coincidence counting for our model system 

{\small{}
\begin{multline*}
{\cal C}\left[\Lambda_{I}\right]\propto,\\
\mathfrak{Re}\sum_{e,e^{\prime}g^{\prime}}\frac{1}{\frac{\omega_{p}}{2}-\omega_{e^{\prime}g^{\prime}}-i\gamma_{e^{\prime}g^{\prime}}}\frac{1}{\frac{\omega_{p}}{2}-\omega_{eg}+i\gamma_{eg}}\left\langle g^{\prime}\vert\left[\mathbbm{1}-iG\left(T\right)\right]\vert g^{\prime}\right\rangle 
\end{multline*}
}{\small\par}

\noindent used in the calculations presented in Fig. $\text{\ref{Fig 4}}$.The
short-lived excited-states of the first excited manifold serve as
a prefactor to the relaxation process of the ground state manifold.
The calculation of Eq. $\text{\ref{eq: Diphasing spectrum}}$ in Fig.
$\text{\ref{Fig 4}}$ where obtained on a discretized grid by scanning
$10^{4}$ frequency, and $10^{3}$ points corresponding to $\omega_{p}$
and $T$ respectively. 

\subsubsection*{2D spectra calculation}

The results for the section were obtained by direct implementation
of Eqs. $S22-S25$ in Sec. $S4$ of the SM. The JSA was discretized
within the shown interval in Fig. $\text{\ref{Fig 5}}$ in a $200\times200\times200\times200$
corresponding to $\omega_{a},\omega_{b},\omega_{p}$ and the implementation
of the numerical integration. The signal shown Fig. $\text{\ref{Fig 5}}$
is obtained by integration over $\omega_{p}$. The ultrashort pump
induces large exchange asymmetry in addition to its broad frequency
range coverage.

\section*{supplementary materials}

Supplementary material for this article are attached.

\bibliographystyle{ScienceAdvances}
\bibliography{EPC}

\begin{thebibliography}{10}

\bibitem{Raymer_2013}
M.~G. Raymer, A.~H. Marcus, J.~R. Widom, D.~L.~P. Vitullo, Entangled
  photon-pair two-dimensional fluorescence spectroscopy (epp-2dfs).
\newblock {\it The Journal of Physical Chemistry B\/} {\bf 117}, 15559-15575
  (2013).

\bibitem{Lavoie_2020}
J.~Lavoie, T.~Landes, A.~Tamimi, B.~J. Smith, A.~H. Marcus, M.~G. Raymer,
  Phase-modulated interferometry, spectroscopy, and refractometry using
  entangled photon pairs.
\newblock {\it Advanced Quantum Technologies\/} {\bf 3}, 1900114 (2020).

\bibitem{Asban_2021}
S.~Asban, K.~E. Dorfman, S.~Mukamel, Interferometric-spectroscopy with
  quantum-light; revealing out-of-time-ordering correlators (2021).

\bibitem{Dorfman_2021}
K.~E. Dorfman, S.~Asban, B.~Gu, S.~Mukamel, Hong-ou-mandel interferometry and
  spectroscopy using entangled photons.
\newblock {\it Communications Physics\/} {\bf 4}, 49 (2021).

\bibitem{Kushing_2020}
S.~Szoke, H.~Liu, B.~P. Hickam, M.~He, S.~K. Cushing, Entangled light--matter
  interactions and spectroscopy.
\newblock {\it Journal of Materials Chemistry C\/} {\bf 8}, 10732-10741 (2020).

\bibitem{Mukamel_2020}
S.~Mukamel, M.~Freyberger, W.~Schleich, M.~Bellini, A.~Zavatta, G.~Leuchs,
  C.~Silberhorn, R.~W. Boyd, L.~L. S{\'a}nchez-Soto, A.~Stefanov, {\it
  et~al.\/}, Roadmap on quantum light spectroscopy.
\newblock {\it Journal of Physics B: Atomic, Molecular and Optical Physics\/}
  {\bf 53}, 072002 (2020).

\bibitem{Mandel_1985}
C.~K. Hong, L.~Mandel, {Theory of parametric frequency down conversion of
  light}.
\newblock {\it Phys. Rev. A\/} {\bf 31}, 2409--2418 (1985).

\bibitem{Bocquillon_2013}
E.~Bocquillon, V.~Freulon, J.-M. Berroir, P.~Degiovanni, B.~Pla{\c c}ais,
  A.~Cavanna, Y.~Jin, G.~F{\`e}ve, Coherence and indistinguishability of single
  electrons emitted by independent sources.
\newblock {\it Science\/} {\bf 339}, 1054--1057 (2013).

\bibitem{Chamon_1997}
C.~de~C.~Chamon, D.~E. Freed, S.~A. Kivelson, S.~L. Sondhi, X.~G. Wen, Two
  point-contact interferometer for quantum hall systems.
\newblock {\it Phys. Rev. B\/} {\bf 55}, 2331--2343 (1997).

\bibitem{Deprez_2021}
C.~D{\'e}prez, L.~Veyrat, H.~Vignaud, G.~Nayak, K.~Watanabe, T.~Taniguchi,
  F.~Gay, H.~Sellier, B.~Sac{\'e}p{\'e}, A tunable fabry--p{\'e}rot quantum
  hall interferometer in graphene.
\newblock {\it Nature Nanotechnology\/}  (2021).

\bibitem{Branning_1999}
D.~Branning, W.~P. Grice, R.~Erdmann, I.~A. Walmsley, Engineering the
  indistinguishability and entanglement of two photons.
\newblock {\it Phys. Rev. Lett.\/} {\bf 83}, 955--958 (1999).

\bibitem{Davydov_1964}
A.~S. Davydov, {THE} {THEORY} {OF} {MOLECULAR} {EXCITONS}.
\newblock {\it Soviet Physics Uspekhi\/} {\bf 7}, 145--178 (1964).

\bibitem{Khalil_2001}
M.~Khalil, A.~Tokmakoff, Signatures of vibrational interactions in coherent
  two-dimensional infrared spectroscopy.
\newblock {\it Chemical Physics\/} {\bf 266}, 213-230 (2001).

\bibitem{Law_2000}
C.~K. Law, I.~A. Walmsley, J.~H. Eberly, Continuous frequency entanglement:
  Effective finite hilbert space and entropy control.
\newblock {\it Phys. Rev. Lett.\/} {\bf 84}, 5304--5307 (2000).

\bibitem{Grice_1997}
W.~P. Grice, I.~A. Walmsley, Spectral information and distinguishability in
  type-ii down-conversion with a broadband pump.
\newblock {\it Phys. Rev. A\/} {\bf 56}, 1627--1634 (1997).

\bibitem{Schlawin_2013b}
F.~Schlawin, S.~Mukamel, Photon statistics of intense entangled photon pulses.
\newblock {\it Journal of Physics B: Atomic, Molecular and Optical Physics\/}
  {\bf 46}, 175502 (2013).

\bibitem{Bahabad_2010}
A.~Bahabad, M.~M. Murnane, H.~C. Kapteyn, Quasi-phase-matching of momentum and
  energy in nonlinear optical processes.
\newblock {\it Nature Photonics\/} {\bf 4}, 570-575 (2010).

\bibitem{Yurke_1986}
B.~Yurke, S.~L. McCall, J.~R. Klauder, Su(2) and su(1,1) interferometers.
\newblock {\it Phys. Rev. A\/} {\bf 33}, 4033--4054 (1986).

\bibitem{Jauche_1976}
F.~R. J.~M.~Jauch, {\it { The Theory of Photons and Electrons}\/} (Springer,
  Berlin, Heidelberg, 1976).

\bibitem{Mota_2004}
R.~D. Mota, M.~A. Xicot{\'{e}}ncatl, V.~D. Granados,
  Jordan{\textendash}schwinger map, 3d harmonic oscillator constants of motion,
  and classical and quantum parameters characterizing electromagnetic wave
  polarization.
\newblock {\it Journal of Physics A: Mathematical and General\/} {\bf 37},
  2835--2842 (2004).

\bibitem{Mota_2004_2}
R.~D. Mota, M.~A. Xicoténcatl, V.~D. Granados, Two-dimensional isotropic
  harmonic oscillator approach to classical and quantum stokes parameters.
\newblock {\it Canadian Journal of Physics\/} {\bf 82}, 767-773 (2004).

\bibitem{Mota_2016}
R.~D. Mota, D.~Ojeda-Guill\'{e}n, M.~Salazar-Ram\'{i}rez, V.~D. Granados,
  Su(1,1) approach to stokes parameters and the theory of light polarization.
\newblock {\it J. Opt. Soc. Am. B\/} {\bf 33}, 1696--1701 (2016).

\bibitem{Mukamel_2008}
S.~Mukamel, Partially-time-ordered schwinger-keldysh loop expansion of coherent
  nonlinear optical susceptibilities.
\newblock {\it Phys. Rev. A\/} {\bf 77}, 023801 (2008).

\bibitem{Dorfman_2014}
K.~E. Dorfman, S.~Mukamel, Multidimensional spectroscopy with entangled light:
  loop vs ladder delay scanning protocols.
\newblock {\it New Journal of Physics\/} {\bf 16}, 033013 (2014).

\bibitem{Schlawin_2013}
F.~Schlawin, S.~Mukamel, Two-photon spectroscopy of excitons with entangled
  photons.
\newblock {\it The Journal of Chemical Physics\/} {\bf 139}, 244110 (2013).

\bibitem{Dorfman_2014b}
K.~E. Dorfman, F.~Schlawin, S.~Mukamel, Stimulated raman spectroscopy with
  entangled light: Enhanced resolution and pathway selection.
\newblock {\it The Journal of Physical Chemistry Letters\/} {\bf 5}, 2843-2849
  (2014).

\bibitem{Hong_1987}
C.~K. Hong, Z.~Y. Ou, L.~Mandel, Measurement of subpicosecond time intervals
  between two photons by interference.
\newblock {\it Phys. Rev. Lett.\/} {\bf 59}, 2044--2046 (1987).

\bibitem{Limonov_2017}
M.~F. Limonov, M.~V. Rybin, A.~N. Poddubny, Y.~S. Kivshar, Fano resonances in
  photonics.
\newblock {\it Nature Photonics\/} {\bf 11}, 543-554 (2017).

\bibitem{Branning_2000}
D.~Branning, W.~Grice, R.~Erdmann, I.~A. Walmsley, Interferometric technique
  for engineering indistinguishability and entanglement of photon pairs.
\newblock {\it Phys. Rev. A\/} {\bf 62}, 013814 (2000).

\bibitem{Mukamel_1995}
S.~Mukamel, {\it {Principles of Nonlinear Optical Spectroscopy}\/} (Oxford
  University Press, 1995).

\bibitem{Marx_2008}
C.~A. Marx, U.~Harbola, S.~Mukamel, Nonlinear optical spectroscopy of single,
  few, and many molecules: Nonequilibrium green's function qed approach.
\newblock {\it Phys. Rev. A\/} {\bf 77}, 022110 (2008).

\bibitem{Helstrom_1976}
H.~C. W., {\it {Quantum Detection and Estimation Theory}\/} (Elsevier, Academic
  Press, Cambridge, Massachusetts, United States, 1976).

\bibitem{Giovannetti_2004}
V.~Giovannetti, S.~Lloyd, L.~Maccone, Quantum-enhanced measurements: Beating
  the standard quantum limit.
\newblock {\it Science\/} {\bf 306}, 1330--1336 (2004).

\bibitem{Giovannetti_review_2011}
V.~Giovannetti, S.~Lloyd, L.~Maccone, Advances in quantum metrology.
\newblock {\it Nature Photonics\/} {\bf 5}, 222-229 (2011).

\bibitem{Napolitano_2010}
M.~Napolitano, M.~W. Mitchell, Nonlinear metrology with a quantum interface.
\newblock {\it New Journal of Physics\/} {\bf 12}, 093016 (2010).

\bibitem{Napolitano_2011}
M.~Napolitano, M.~Koschorreck, B.~Dubost, N.~Behbood, R.~J. Sewell, M.~W.
  Mitchell, Interaction-based quantum metrology showing scaling beyond the
  heisenberg limit.
\newblock {\it Nature\/} {\bf 471}, 486-489 (2011).

\bibitem{HOWELLS_2009}
M.~Howells, T.~Beetz, H.~Chapman, C.~Cui, J.~Holton, C.~Jacobsen, J.~Kirz,
  E.~Lima, S.~Marchesini, H.~Miao, D.~Sayre, D.~Shapiro, J.~Spence,
  D.~Starodub, An assessment of the resolution limitation due to
  radiation-damage in x-ray diffraction microscopy.
\newblock {\it Journal of Electron Spectroscopy and Related Phenomena\/} {\bf
  170}, 4 - 12 (2009). Radiation Damage.

\bibitem{Roslyak_2009}
O.~Roslyak, C.~A. Marx, S.~Mukamel, Nonlinear spectroscopy with entangled
  photons: Manipulating quantum pathways of matter.
\newblock {\it Phys. Rev. A\/} {\bf 79}, 033832 (2009).

\bibitem{Dorfman_rev_2016}
K.~E. Dorfman, F.~Schlawin, S.~Mukamel, Nonlinear optical signals and
  spectroscopy with quantum light.
\newblock {\it Rev. Mod. Phys.\/} {\bf 88}, 045008 (2016).

\bibitem{Kimble_1977}
H.~J. Kimble, M.~Dagenais, L.~Mandel, Photon antibunching in resonance
  fluorescence.
\newblock {\it Phys. Rev. Lett.\/} {\bf 39}, 691--695 (1977).

\bibitem{Hell_1994}
S.~W. Hell, J.~Wichmann, Breaking the diffraction resolution limit by
  stimulated emission: stimulated-emission-depletion fluorescence microscopy.
\newblock {\it Opt. Lett.\/} {\bf 19}, 780--782 (1994).

\bibitem{Mouradian_2011}
S.~Mouradian, F.~N.~C. Wong, J.~H. Shapiro, Achieving sub-rayleigh resolution
  via thresholding.
\newblock {\it Opt. Express\/} {\bf 19}, 5480--5488 (2011).

\bibitem{Schwartz2013}
O.~Schwartz, J.~M. Levitt, R.~Tenne, S.~Itzhakov, Z.~Deutsch, D.~Oron,
  Superresolution microscopy with quantum emitters.
\newblock {\it Nano Letters\/} {\bf 13}, 5832-5836 (2013).

\bibitem{Tenne_2019}
R.~Tenne, U.~Rossman, B.~Rephael, Y.~Israel, A.~Krupinski-Ptaszek,
  R.~Lapkiewicz, Y.~Silberberg, D.~Oron, Super-resolution enhancement by
  quantum image scanning microscopy.
\newblock {\it Nature Photonics\/} {\bf 13}, 116-122 (2019).

\bibitem{Javanainen_1990}
J.~Javanainen, P.~L. Gould, Linear intensity dependence of a two-photon
  transition rate.
\newblock {\it Phys. Rev. A\/} {\bf 41}, 5088--5091 (1990).

\bibitem{Lee_2006}
D.-I. Lee, T.~Goodson, Entangled photon absorption in an organic porphyrin
  dendrimer.
\newblock {\it The Journal of Physical Chemistry B\/} {\bf 110}, 25582-25585
  (2006).

\bibitem{Guzman_2010}
A.~R. Guzman, M.~R. Harpham, {\"O}.~S{\"u}zer, M.~M. Haley, T.~G. Goodson,
  Spatial control of entangled two-photon absorption with organic chromophores.
\newblock {\it Journal of the American Chemical Society\/} {\bf 132}, 7840-7841
  (2010).

\bibitem{Lantz_2014}
E.~Lantz, P.-A. Moreau, F.~Devaux, Optimizing the signal-to-noise ratio in the
  measurement of photon pairs with detector arrays.
\newblock {\it Phys. Rev. A\/} {\bf 90}, 063811 (2014).

\bibitem{Bolduc_2017}
E.~Bolduc, D.~Faccio, J.~Leach, Acquisition of multiple photon pairs with an
  {EMCCD} camera.
\newblock {\it Journal of Optics\/} {\bf 19}, 054006 (2017).

\bibitem{Reichert_2018}
M.~Reichert, H.~Defienne, J.~W. Fleischer, Massively parallel coincidence
  counting of high-dimensional entangled states.
\newblock {\it Scientific Reports\/} {\bf 8}, 7925 (2018).

\bibitem{Asban_2019}
S.~Asban, S.~Mukamel, Scattering-based geometric shaping of photon-photon
  interactions.
\newblock {\it Phys. Rev. Lett.\/} {\bf 123}, 260502 (2019).

\bibitem{Asban_2021b}
S.~Asban, F.~J. Garc{\'i}a~de Abajo, Generation, characterization, and
  manipulation of quantum correlations in electron beams.
\newblock {\it npj Quantum Information\/} {\bf 7}, 42 (2021).

\end{thebibliography}

\noindent \textbf{Acknowledgments:} The support of the National Science
Foundation (NSF) Grant CHE-1953045 is gratefully acknowledged.\textbf{
Author contributions:} Both authors have contributed equally to this
work.\textbf{ Competing interests:} The authors declare no competing
interests\textbf{. Data and materials availability:} The main results
of this manuscript are composed of analytical and numerical calculations.
All data generated, analyzed or required to reproduce the results
of this study are included in this article and its Supplementary Material
file.
\end{document}